\documentclass[12pt]{article}
\pdfoutput=1 
\usepackage[margin=1.2in]{geometry}
\usepackage{authblk}
\usepackage[font=small]{caption}

\usepackage{graphicx}
\usepackage{amsmath}
\usepackage{amssymb}
\usepackage{hyperref}

\usepackage{setspace}
\title{\bf \large Self-Replicating Hierarchical Structures Emerge in a Binary Cellular Automaton}

\author{\normalsize Bo Yang\\
\small Adjacent Lab, New York, United States\\
\small bo@adjacentlab.com }

\date{}

\begin{document}

\maketitle

\begin{abstract}
We have discovered a novel transition rule for binary cellular automata (CA) that yields self-replicating structures across two spatial and temporal scales from sparsely populated random initial conditions. Lower-level, shapeshifting clusters frequently follow a transient attractor trajectory, generating new clusters, some of which periodically self-duplicate. When the initial distribution of live cells is sufficiently sparse, these clusters coalesce into larger formations that also self-replicate. These formations may further form the boundaries of an expanding complex on an even larger scale. This rule, dubbed ``Outlier,'' is rotationally symmetric and applies to 2D Moore neighborhoods. It was evolved through Genetic Programming during an extensive automated search for rules that foster open-ended evolution in CA. While self-replicating structures, both crafted and emergent, have been created in CA with state sets intentionally designed for this purpose, the Outlier may be the first known rule to facilitate emergent self-replication across two spatial scales in simple binary CA.
\end{abstract}

\section{Background and Introduction}

Self-replication, a hallmark of biological life, represents a significant milestone in the pursuit of artificial life across various mediums. In the digital realm, Von Neumann, in tandem with the earliest cellular automaton, envisioned self-replicating machines as a pathway towards universal construction \cite{vonNeumann1966}. His automaton involved cells transitioning among 29 states, and the initial structure spanned hundreds of thousands of cells, necessitating meticulous design. As simpler cellular automata (CA) demonstrated their unique utility, self-replication—regardless of potential applications in universal construction—emerged as a distinct area of interest. Most notably, Langton constructed an eight-state automaton with an 86-cell initial loop structure capable of self-replication \cite{langton1984}. Subsequent research has sought to simplify this further or enhance its capabilities \cite{sayama1999}.

Traditionally, most self-replicating structures were manually designed. However, CA capable of forming self-replicating structures from random initial conditions could expand our understanding of self-organization and emergence. In \cite{chou1997}, this was accomplished using CA with 8-bit state sets, each segmented into four parts to facilitate elements of the replication process, such as signal flow, movement, bonding, and detachment. Self-replicating loops, composed of 2 by 2 cells, emerge from random initial conditions, and their sizes may increase over time.

\begin{figure}[!b]
    \centering
    \includegraphics[width=\linewidth]{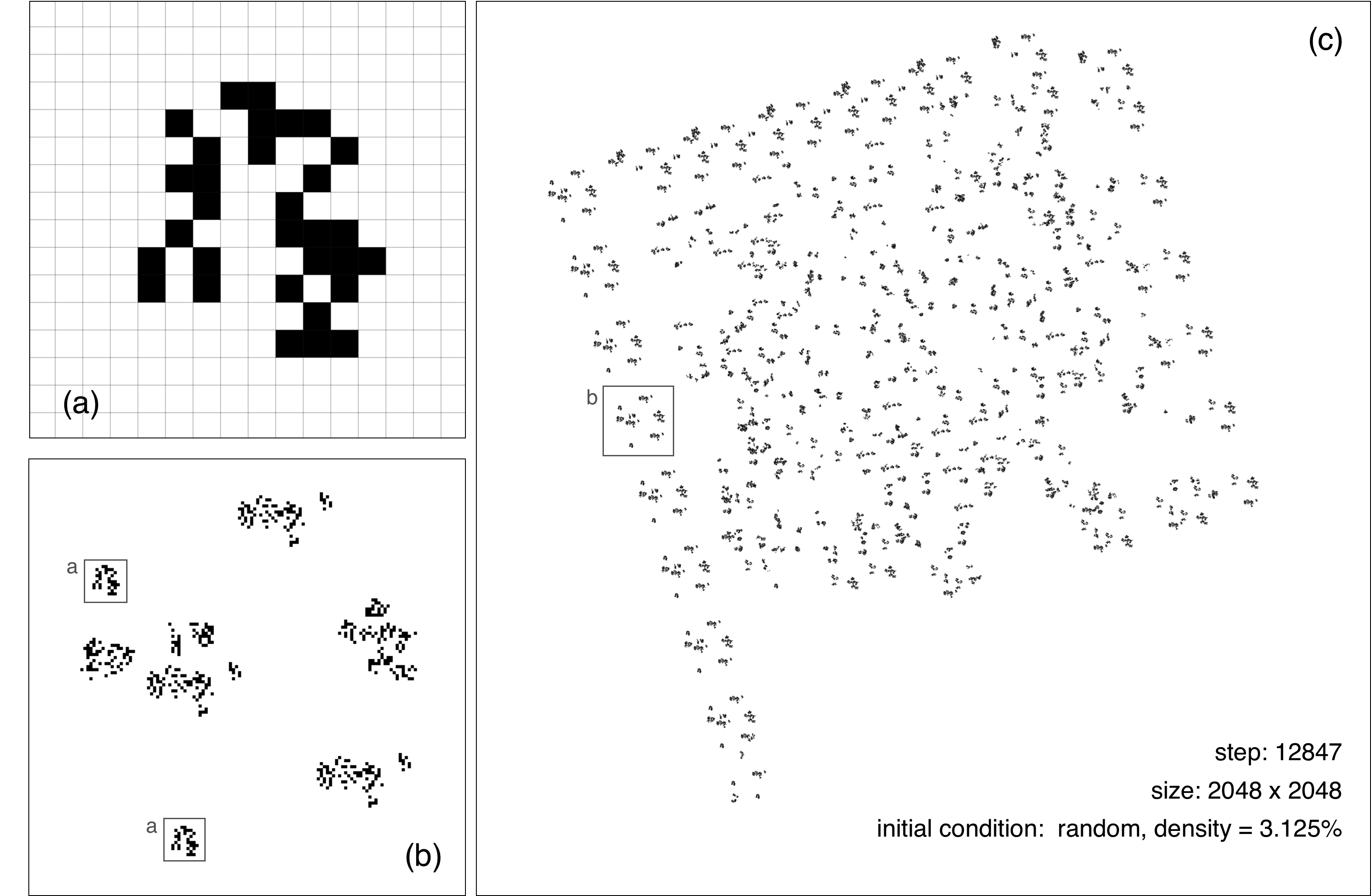}
    \caption{Sample outcome from the Outlier rule starting with a sparse random initial condition. (a) Two clusters on the smallest scale; (b) A self-replicating formation, assembled from a few clusters; (c) On the largest scale, an expanding complex with a semi-chaotic interior, bordered by replicating formations.}
    \label{fig:complex}
\end{figure}

In this paper, we report the discovery of a novel two-state CA rule that enables the spontaneous assembly of larger self-replicating ``formations'' from smaller, shapeshifting ``clusters'' that themselves emerge from random initial conditions. An increasing number of these replicating formations often subsequently form an expanding superstructure, or ``complex,'' on an even larger scale. Figure \ref{fig:complex} illustrates the hierarchical arrangement of these structures. We have named this rule the ``Outlier,'' as it generates the most seemingly complex behaviors among all the interesting rules we have encountered.

\section{Discovery of the Outlier Rule}

The Outlier rule was serendipitously discovered during an extensive automated search for CA rules that would support open-ended evolution (OEE), defined as the continuous emergence of novel and increasingly complex behaviors ~\cite{bedau2000open}. Although often associated with OEE, self-replication was never an explicit search goal in that study. Consequently, we will defer most implementation details to a future report, focusing here on two aspects of the methods relevant to the characteristics of the rules selected for evaluations.

The particular search runs that led to the Outlier were conducted within the space of $2^{140}$ rotationally symmetric rules on Moore neighborhoods of 2D binary CA, $\mathcal{R}$. Mirror parity was not required. To render the search tractable, we employed Genetic Algorithm (GA), Genetic Programming (GP), and various forms of bit representation of the rules as genotypes in several phases of the project.

In general, the details of genotype representation in GA and GP searches modulate the probability distribution of random sampling in the parameter space, thereby shaping the search path. This fact is particularly significant in our project, as the entirety of all rules ever evaluated will comprise an infinitesimally small fraction of $\mathcal{R}$. The Outlier rule was discovered during a Genetic Programming search run wherein each genotype, or rule, was represented as a tree structure of bitwise logic operations, $(G_1, \dots, G_L)$. Specifically, each node $G_i$ in the tree of length $L$ is a tuple of three integers:

\begin{align*}
&G_i = (f_i, i_1, i_2) \text{ where } i_1 <= i_2 < i <= L, \text{ and } f_i = 0 \text{ or } 1\
\intertext{For each cell in a Moore neighborhood, we then compute its new state by traversing the entire tree, starting from its neighborhood states:}
&N_i =
\begin{cases}
N_{i_1} \oplus N_{i_2}, & \text{if}\ f_i = 1 \\
N_{i_1} \wedge N_{i_2}, & \text{if}\ f_i = 0
\end{cases} \text{ for } i > 9, \\
&N_i = S_i , \text{ for } i \le 9
\end{align*}

Here, $S_1$ through $S_9$ represent the current states of the cells in the Moore neighborhood, with $S_1$ at the center. The center cell is then updated to $N_L$ for the next step, along with all other cells in the automaton updated in the same manner. Additional procedures were added before and after tree traversals to enforce rotational symmetry. The traditional lookup table representation of each rule can be mapped to several logical trees expressed in this way, and they are computationally equivalent.

The choice of this representation was initially motivated by computational efficiency, crucial to CA rule search, as a single fitness evaluation often necessitates the calculation of hundreds of billions of cell updates. With one bit per cell memory representation, many adjacent cells can be loaded into long word registers and updated in parallel via consecutive bitwise logical operations specified by the aforementioned trees. Modern CPUs and GPUs can update hundreds to billions of bits concurrently in this manner, with excellent memory locality. For instance, the GP search in 2019 that resulted in the Outlier rule ran on a 14-core Xeon CPU, capable of updating thousands of cells concurrently with AVX-512 support in each core. In later iterations on GPUs, bitwise logical operation trees were tweaked to keep most, if not all, operations in the GPU register files, thanks to optimizing kernel compilers. This often resulted in a speedup by two orders of magnitude.

The second implementation detail pertinent to our findings is the fitness function, which measures the complexity or ``open-endedness'' of the phenotypes, which in our case are the CA bitmaps generated by each rule. As often happens in GA/GP searches, fitness functions derived directly from spatial and temporal analysis are prone to ``cheating,'' wherein rules maximize the fitness score with surprisingly novel yet simplistic behaviors. In the later stages of the project, we adopted ``novelty search'' as first developed by Lehman and Stanley ~\cite{lehman2008}. This approach rewards new phenotypes that brings ``novelty'' to all previously evaluated phenotypes. In our implementation, we extract a feature vector, $\mathbf{F}$, for each rule by quantifying the complexity of CA bitmaps in the later stages of convergence. For each new rule, a novelty score is calculated from the distances from $\mathbf{F}$ to its $k$ nearest neighbors in the space of all (or a large sample) of previously computed $\mathbf{F}$.

This implementation of novelty search was somewhat successful, yielding a few rules with intriguing behaviors not seen with other fitness functions. This includes the Outlier, which was algorithmically tagged as sufficiently ``novel'' for visual inspection. However, nothing substantially more complex has been observed thus far, and our search for OEE in CA continues.

\begin{figure}[!t]
\centering
\includegraphics[width=\linewidth]{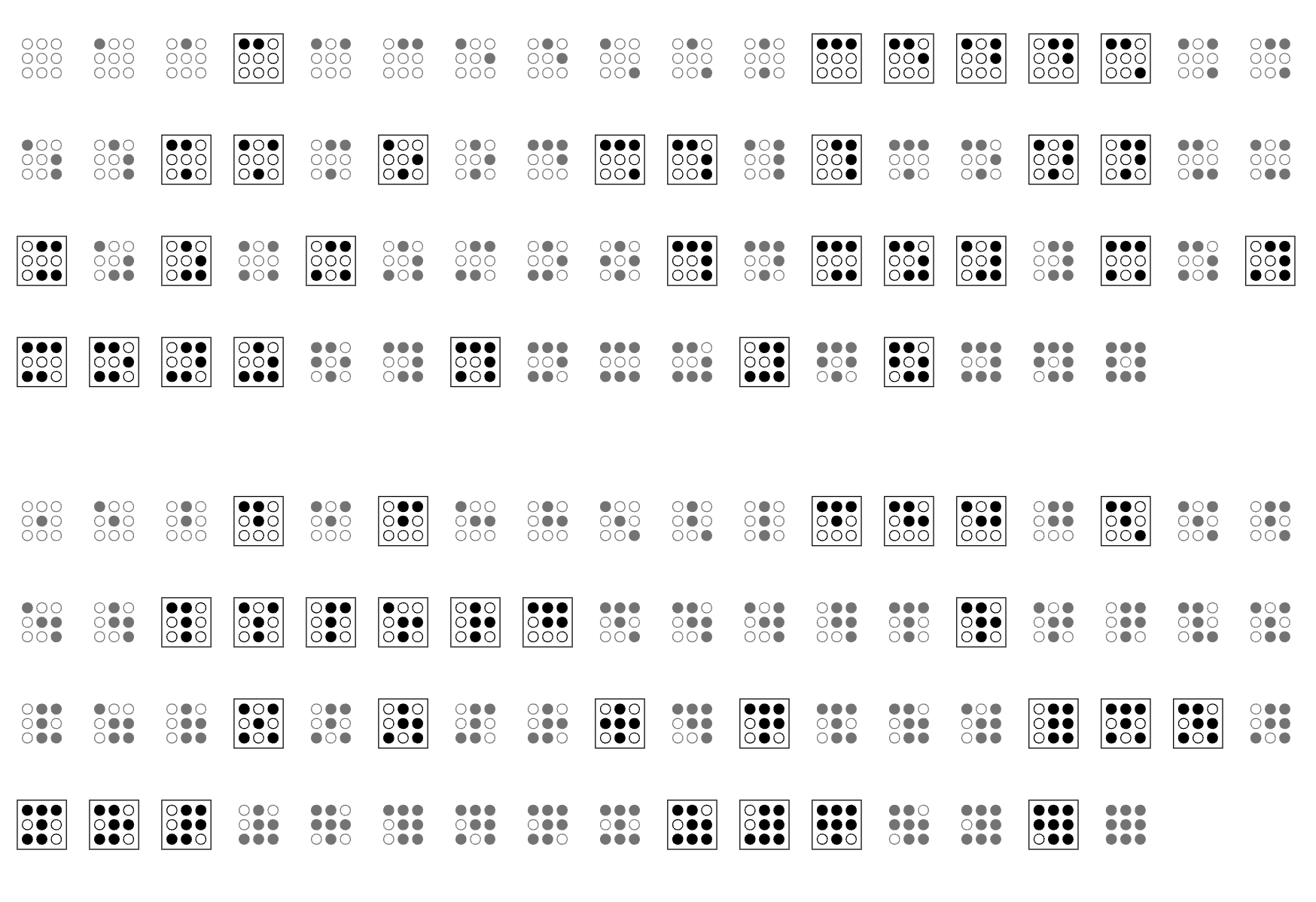}
\caption{The Outlier Rule. The center cell in each of the boxed neighborhoods and their three quarter-turn rotations stays alive. Filled/empty circles stand for live/dead cells respectively. }
\label{fig:outlier}
\end{figure}

\section{Cluster, Formation, and Complex}

As listed in Figure \ref{fig:outlier}, the Outlier rule observes rotational symmetry but lacks mirror symmetry. Similar to many solutions produced by GA/GP, it does not possess a clearly recognizable structure or definable formulation. Notably, its rule table representation has 220 live entries out of 512, which is denser than Conway's Game of Life, which has 140.

Under this rule, three categories of trajectories typically follow random initial configurations. Although each individual outcome is probabilistic, the statistical likelihoods are highly dependent on the initial density of live cells, $D_0$, and the grid size. A 1024 by 1024 grid is more likely to become completely empty when $D_0 < 0.02$, semi-chaotic when $D_0 > 0.15$, and likely to support replicating formations when $D_0$ falls in between these values. We will refer to these three types of outcomes as ``barren,'' ``dense,'' and ``sparse,'' respectively. These cutoff values for $D_0$ are grid-size dependent. For grids smaller than 512 by 512, replicating formations do not occur at all. We will explain the dependency of the likelihoods on $D_0$ in the next section.

\begin{figure}[!b]
\centering
\includegraphics[width=\linewidth]{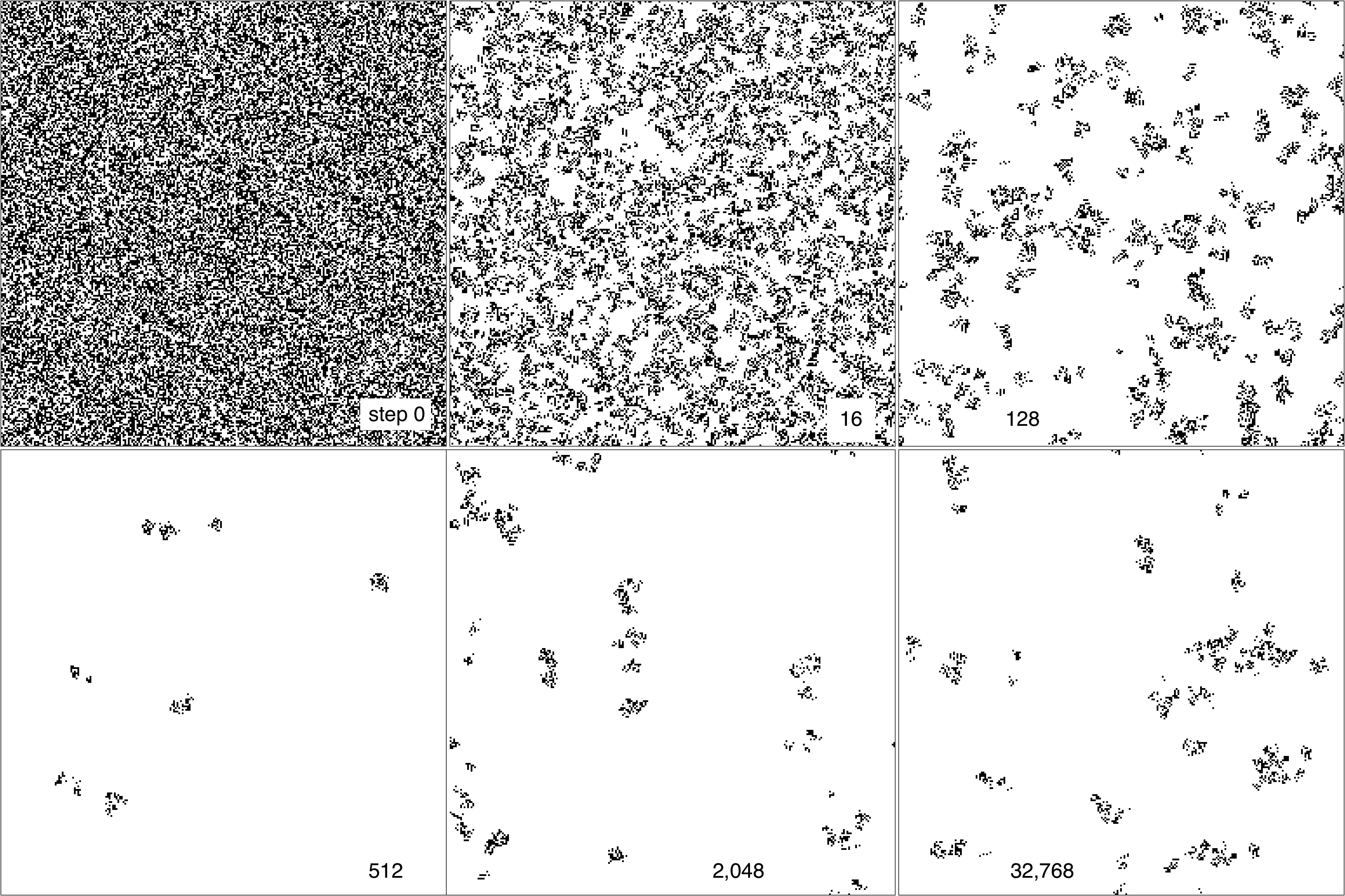}
\caption{A densely populated random grid transitions to a semi-chaotic phase. The grid is 256 by 256 cells, with periodic boundary conditions, and an initial density of 50\%. The step numbers are shown as labeled.}
\label{fig:dense}
\end{figure}

Regardless of $D_0$, shape-shifting clusters, each composed of a few dozen live cells at most, form in fewer than a hundred steps. A ``cluster'' is formally characterized as evolving shapes composed of live cells that are topologically connected; two live cells are considered adjacent if they are in each other's Moore neighborhood. Each cluster continuously shape-shifts, sometimes splitting into two, or interacting with another cluster through collision or merging. Without these interactions, most of these clusters would disappear within a hundred steps.

However, on a ``sparse'' grid that is sufficiently large, a small fraction of the clusters can survive and grow into larger self-replicating formations by spawning new clusters. Each of these formations consists of groups of clusters, with the numbers fluctuating around ten. A replicating formation expands its territory by creating copies of itself while slowly traversing, until it collides with another formation or cluster outside its territory. Collisions break down a formation back to clusters, which then change shape and interact among themselves continuously and in a chaotic manner, eventually occupying the entire grid. As shown in Figure \ref{fig:dense}, a dense grid will transition directly into this semi-chaotic phase, before any formation has the opportunity to emerge.

When a single replicating formation survives in the middle of a sufficiently large empty area, it periodically generates new formations that form an even larger structure, or a ``complex,'' at a still higher scale. Each boundary region of a complex's four edges consists mostly of replicating formations. These are mostly identical to each other and shape-shift synchronously. The initial expansion of the boundaries appears to be driven by a formation protruding out of the rectangular boundary, as shown in Figure \ref{fig:complex}. Given that all replicating formations are spaced (52, 172) or (172, 52) cells apart from neighboring formations on each side of the complex, the edges of a complex form a rectangle that is tilted counter-clockwise from the axes by $\arctan(13/43)$, or approximately 16.8214 degrees.

The interior of a complex is occupied by surplus clusters, or ``debris,'' that are generated by the replication process but are not part of the replicating formation themselves. These evolve in the same semi-chaotic manner as on a dense grid, and interact with the bordering formations without affecting the integrity of the formations. A complex continuously expands until it occupies all available space or until it collides with other structures outside its territory. Upon such collision, formations break down, and their clusters continue their dynamic transformations on a lower scale, similar to a dense grid.

\section{Temporal Loops and Transient Attractors}

\begin{figure}[!b]
\centering
\includegraphics[width=\linewidth]{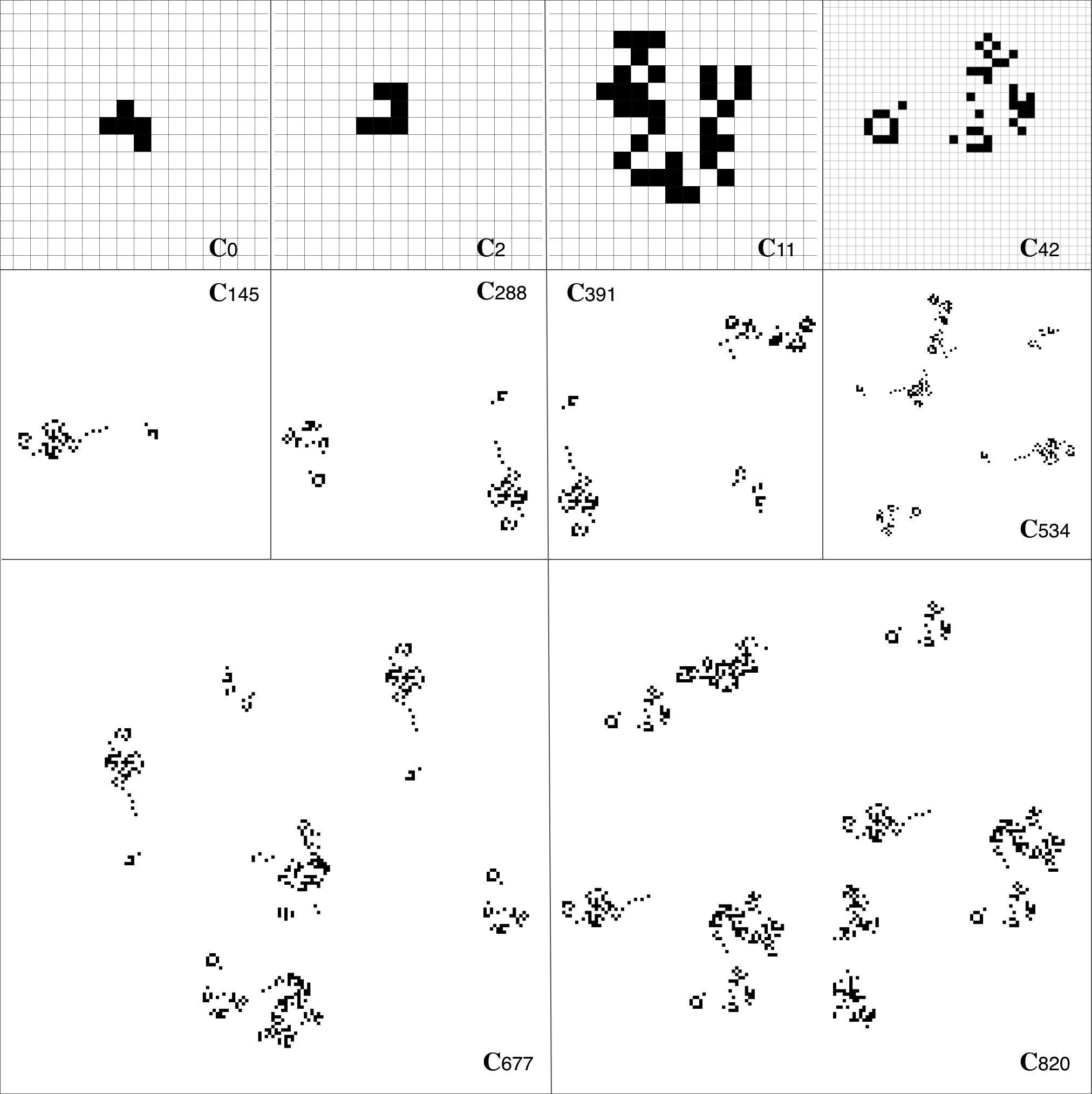}
\caption{Seed Trajectory $\mathcal{A}s$. Configurations are numbered with step counts. $\mathbf{c}2$ reappears periodically every 143 steps in $\mathbf{c}2$ - $\mathbf{c}{145}$ - $\mathbf{c}{288}$, and $\mathbf{c}{391}$ - $\mathbf{c}{534}$ - $\mathbf{c}{677}$. The formation in $\mathbf{c}{391}$ is already capable of self-replication, as shown in the next figure. Note rotations of $\mathbf{c}{11}$ appears in Figure \ref{fig:complex}.}
\label{fig:seed}
\end{figure}

To understand the dynamics underlying the replicating formations, we conducted experiments by initializing an empty grid with a single isolated 3 by 3 cluster in the center. Results are displayed in Figure \ref{fig:seed}. Out of the 140 possible initial configurations, considering rotational symmetry, two ($\mathbf{c}_0$ and $\mathbf{c}_2$) evolve into replicating formations, while all others die out. $\mathbf{c}_0$ updates into $\mathbf{c}_2$ in two steps and thus follows the same trajectory thereafter. We refer to them as ``seed'' clusters and their trajectory as $\mathcal{A}_s$. In short, any isolated 3 by 3 initial cluster either disappears or follows $\mathcal{A}_s$.

A detailed examination of $\mathcal{A}_s$ reveals that $\mathbf{c}_2$ and some of its follow-up clusters reappear periodically, rotated 90 degrees counter-clockwise each time, with a period of 143 steps. The first period begins when $\mathbf{c}_2$ is initialized and ends when it reappears in $\mathbf{c}_{145}$, rotated and translated, amongst several clusters spun off during the period. This happens again after another 143 steps, and the formation grows larger. Another two-period run starts at $\mathbf{c}_{391}$, with two then three rotated copies of $\mathbf{c}_2$. Each new reappearance of the rotated $\mathbf{c}_2$ introduces a new sub-trajectory into $\mathcal{A}_s$ if the new cluster is sufficiently isolated from the rest of the formation and can thus seed its own trajectory. Because the rule is deterministic and rotationally symmetric, all the structures appearing in the first period, such as $\mathbf{c}_{11}$ and $\mathbf{c}_{42}$, reappear and sometime self-replicate in the same 143-step period. We also identify these as "seed clusters," and each time one materializes outside the existing trajectory, it adds a new branch, or sub-trajectory, onto $\mathcal{A}_s$.

These sub-trajectories are only partially self-similar to the original $\mathcal{A}_s$, as collisions restrict their growth when the vicinity becomes crowded. Hence, four-period reappearances rarely occur. For example, in $\mathbf{c}_{820}$, which is 143 steps after $\mathbf{c}_{677}$, four copies of $\mathbf{c}_{42}$ appear instead of $\mathbf{c}_2$. $\mathbf{c}_{391}$ appears to be around the time when the accumulation of new clusters suppresses dynamics of 143-step periods by crowding the empty space, and inter-cluster interactions form parallel dynamics on a longer timescale, embodied in larger structures emerging at the formation scale, some of which can self-replicate and thus be identified visually.

\begin{figure}[!b]
\centering
\includegraphics[width=\linewidth]{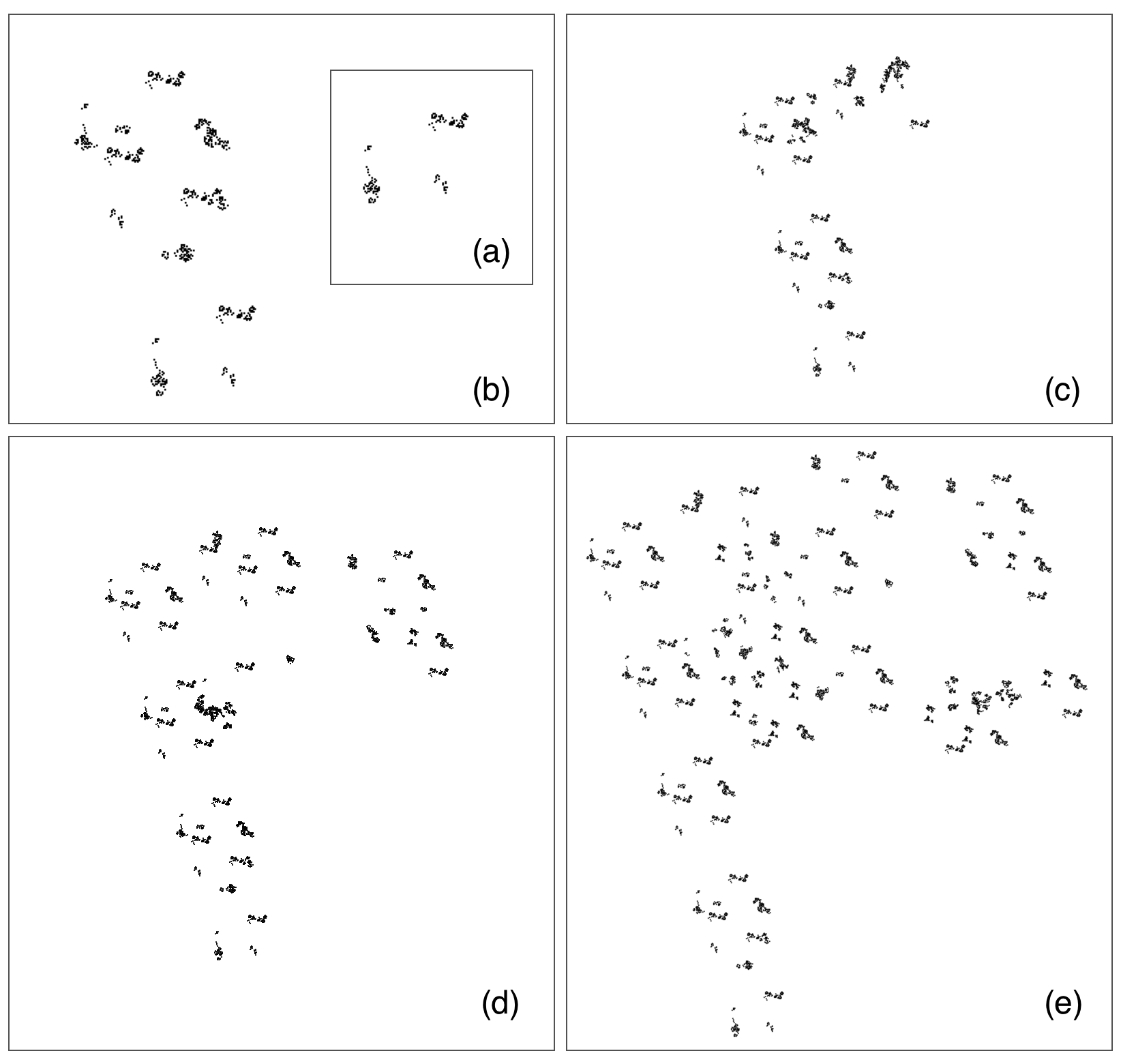}
\caption{Formations self-replicating every 1556 steps. Step counts: (a) 391, (b) 1947, (c) 3503,  (d) 5059, (e) 6615.}
\label{fig:formations}
\end{figure}

The formations in the third 143-step period, for example, the one in Figure \ref{fig:seed} $\mathbf{c}_{391}$, as a whole and as a higher-level structure, start to reappear with a period of 1556 steps. In fact, it evolves into the ``protruding arm'' of the larger complex, as illustrated in Figure \ref{fig:formations}. Visually, it appears to be shape-shifting while slowly moving away from its original position, producing a new replicating formation ``behind'' itself in each period. A closer inspection reveals that it shares many clusters with the adjacent replicating formation that is being formed, and the boundary between the formations shifts constantly and lacks a clear definition. Many identical clusters are components of both the protruding arm and the replicating formations and they shape-shift in sync, and most of these clusters and formations reappear every 1556 steps.

A similar replicating process later starts at the diagonally opposite corner of the complex, with the corner formation appearing to be ``caved in'' rather than protruding. In Figure \ref{fig:formations}, the replicating process first starts at the left edge of the complex, then at each following edge in a clockwise order, spaced by a time lag characterized by the same 1556-step period. The bottom edge forms last and is least defined. As the complex expands, one new replicating formation is added to each edge every 1556 steps. On each edge, most of its formations shape-shift in perfect sync and repeat with the same 1556-step period.

Under sparse initial random conditions, our close examination of the updates revealed a clear pattern: the majority of arbitrarily formed clusters eventually vanish. The rare survivors consistently follow the same trajectory $\mathcal{A}_s$, aside from a small period-four "spinner" that rotates 90 degrees per step, usually getting absorbed by other clusters. For example, the automaton in Figure \ref{fig:complex} has one surviving cluster that enters $\mathcal{A}_s$ as a seed cluster $\mathbf{c}_2$. When more than one cluster survive, their individual evolutions along $\mathcal{A}_s$ derail when there is a collision between clusters originating from different seeds. Consequently, for the evolution into replicating formations, surviving clusters need to maintain sufficient spatial separation.  This explains how the density of the initial random configuration determines the existential likelihood of replicating formations.

Neither random initial conditions nor 3 by 3 initial seeds cleanly generate ``pure'' replicating formations, as they always produce additional clusters, or ``debris,'' in their vicinities. Out of curiosity, we initialized a grid with nothing but an isolated replicating formation, without the debris, and found the subsequent behavior to be similar, as it self-replicates and then grows into a complex. Additionally, we isolated each individual component cluster in a replicating formation and successively used each as an individual seed for initialization, and found that about half of them disappear, while the others evolve into full formations. In short, $\mathcal{A}_s$ appears to be dominating, even though it is not robust.

In conclusion, a cellular automaton operating under the Outlier rule transitions into one of three phases: empty, semi-chaotic, or replication at the formation level. The last phase is characterized by a trajectory that resembles an expanding transient attractor. Reappearances of both clusters and formations attach sub-loops to the trajectory, even though they have different characterizing period length: 143 steps for clusters, and 1556 steps for formations. However, this attractor is only transient, as eventually the complex runs out of empty space to expand into, or it collides with other structures, and the semi-chaotic phase takes over. But between that eventuality and the initial randomness, replicating formations can exist for quite many steps.

\section{Discussion}

The ``building blocks'' one level down from replicating structures in previously constructed self-replicating CAs are the cells themselves, each in one of multiple states, the number of which ranges from 8 to the hundreds~\cite{langton1984}~\cite{sayama1999}~\cite{chou1997}. Each of these states or their subcomponents was assigned a primary ``role,'' such as information storage, replication trigger, structural protection, collision avoidance or inducement. Similar to components of engineered machinery, a state often plays multiple roles but never all of them. In contrast, emergent self-replication in binary CAs has to be more complex than complicated, as, apart from its rule encodable in slightly more than a hundred bits, everything else must emerge on its own. Each cell in a binary CA carries minimal information. The 'building blocks', therefore, must be clusters of cells that emerge from the rule. In this context, emergence on multiple scales becomes a necessity for self-replication.

However, it is unclear in the case of the Outlier rule, if each of the clusters carries a role that is specific to the assembly of a replicating formation. The clusters seem to be different but among equals, and perhaps they lack such specificity due to the constraints imposed by the shared cell-level updating rule. Instead, processes on a higher scale level emerge from interactions among clusters in proximity, which in return supports the continued existence and evolution of the clusters. We have observed similar inter-cluster dynamics with other CA rules, but the higher level processes always appear chaotic. The Outlier rule is exceptional in that some of its emergent processes self-repeat, embodied in self-replicating formations.

Interestingly, the larger complex generated by the Outlier rule, although not self-replicating as a whole, presents a boundary shape that superficially resembles the ``loop''-shaped self-replicators as designed in ~\cite{langton1984}~\cite{sayama1999} or emerged in ~\cite{chou1997}, specifically, a rectangle with a protruding arm. Whether this resemblance is coincidental or substantial warrants further investigation.

The Outlier is the only rule that can generate replicating formations amongst the few hundred thousand rules we examined. Yet its composition looks irregular and arbitrary, which begs the question of how common similarly capable rules are in $\mathcal{R}$. We performed one-bit flips, or single configuration mutations of the representation in Figure \ref{fig:outlier}, and found no such capability in all the mutations. It appears that the Outlier is unique at least in its immediate adjacent rule space. This of course helps very little in answering the question. Nevertheless, the fact that nontrivial emergent behavior occurs on multiple scales in a simple binary cellular automaton can be intriguing, and this author hopes it is illustrative as well.

Finally, inching towards open-ended evolution in CA, a logical adjacent step would be to identify a rule that supports not only emergent replication but also adaptation and structural evolution and is robust to collisions. Some of such capabilities have already been previously showcased with specially designed states, as seen in the nine-state automaton initialized with loop structures discussed in ~\cite{sayama1999}. The prospect of such rules existing with simpler CA remains uncertain. But certainly, only an infinitesimally small fraction of the vast rule space has ever been explored thus far.

\section{Acknowledgements}
The author would like to thank Bert Chan and Hiroki Sayama for their encouragement to write this down before venturing away further.

\bibliographystyle{unsrt}
\bibliography{references}

\begin{thebibliography}{1}

\bibitem{vonNeumann1966}
John von Neumann.
\newblock {\em Theory of self-reproducing automata}.
\newblock University of Illinois Press, 1966.

\bibitem{langton1984}
Christopher~G Langton.
\newblock Self-reproduction in cellular automata.
\newblock {\em Physica D: Nonlinear Phenomena}, 10(1-2):135--144, 1984.

\bibitem{sayama1999}
Hiroki Sayama.
\newblock A new structurally dissolvable self-reproducing loop evolving in a
  simple cellular automata space.
\newblock {\em Artificial Life}, 5:343--365, 1999.

\bibitem{chou1997}
H.~H. Chou and J.~A. Reggia.
\newblock Emergence of self-replicating structures in a cellular automata
  space.
\newblock {\em Physica D: Nonlinear Phenomena}, 110:252--276, 1997.

\bibitem{bedau2000open}
Mark~A Bedau, John~S McCaskill, Norman~H Packard, Steen Rasmussen, Christoph
  Adami, David~G Green, Takashi Ikegami, Kunihiko Kaneko, and Thomas~S Ray.
\newblock Open problems in artificial life.
\newblock {\em Artificial life}, 6(4):363--376, 2000.

\bibitem{lehman2008}
Joel Lehman and Kenneth~O. Stanley.
\newblock Exploiting open-endedness to solve problems through the search for
  novelty.
\newblock In {\em Proceedings of the Eleventh International Conference on the
  Synthesis and Simulation of Living Systems, {ALIFE} 2008}, pages 329--336.
  {MIT} Press, 2008.

\end{thebibliography}

\end{document}